\documentclass[iop]{emulateapj}

\begin{document}

\title{Predicting $\alpha$ Comae Berenices Time of Eclipse II:\\How 3 Faulty Measurements Out of 609 Caused A 26 Year Binary's Eclipse To Be Missed}

\author{Matthew W.~Muterspaugh\altaffilmark{1, 2}, 
M.~J.~P.~Wijngaarden\altaffilmark{3}, 
H.~F.~Henrichs\altaffilmark{3}, 
Benjamin F.~Lane\altaffilmark{4}, 
William I.~Hartkopf\altaffilmark{5}, 
and Gregory W.~Henry\altaffilmark{2}}
\altaffiltext{1}{Department of Mathematical Sciences, College of Engineering, 
Tennessee State University, Boswell Science Hall, Nashville, TN 
37209 }
\altaffiltext{2}{Center of Excellence in Information Systems, Tennessee State University, 3500 John A. Merritt Blvd., Box No.~9501, Nashville, TN 
37209-1561}
\altaffiltext{3}{Anton Pannekoek Institute for Astronomy, University of Amsterdam, Science Park 904, 1098 XH Amsterdam, Netherlands}
\altaffiltext{4}{Draper Laboratory,  555 Technology Square, Cambridge, MA 
02139-3563}
\altaffiltext{5}{U.S.~Naval Observatory, 3450 Massachusetts Avenue, NW, Washington, DC, 20392-5420}

\email{matthew1@coe.tsuniv.edu, gregory.w.henry@gmail.com}

\begin{abstract}
The dwarf stars in the 26 year period binary $\alpha$ Com were predicted to 
eclipse each other in early 2015.  That prediction was based on an orbit model 
made with over 600 astrometric observations using micrometers, speckle 
interferometry, and long baseline optical interferometry.  Unfortunately, it 
has been realized recently that the position angle measurements for three of 
the observations from $\sim 100$ years ago were in error by 180 degrees, which 
skewed the orbital fit.  The eclipse was likely 2 months earlier than 
predicted, at which point the system was low on the horizon at sunrise.
\end{abstract}

\section{Introduction}

\begin{figure}[]
\epsscale{1.2}
\plotone{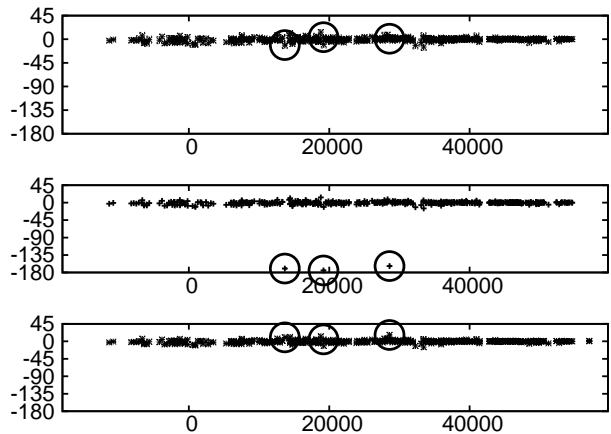}
\caption[HD 114378 ($\alpha$ Com) Orbit Fit Residuals]
{ \label{fig::114378_resid}
Fit residuals (observed-computed) for the positional angles of $\alpha$ Com for various orbit models and 180 degree ambiguities.  The horizontal axis is modified julian day and the vertical axis is residual angle in degrees.  In each case, the 3 problematic measurements from 1896.33, 1911.4, and 1937.16 are circled.
(Top)  Fit residuals as the data were presented for the model by \cite{2010AJ....140.1623M}.  Even though the three measurements were input with 180 degree errors, the model fitting procedure resulted in a model skewed enough that the measurements do not appear as obvious outliers.  (Middle)  Fit residuals for a model fit using only the separation measurements and PHASES observations, then applied to the positional angle data.  The problem observations are clear.  (Bottom)  After fixing the 180 degree ambiguities of the three measurements, residuals to the currently presented orbital model.  Measurements from sources other than the PHASES program and CHARA array are from the WDS \citep{wdsCatalog}.
}
\end{figure}

$\alpha$ Comae Berenices has long been suspected of eclipsing \citep{1875MNRAS..35..367S, Hart1989, hoffleit1996} despite being a 26 year binary, due to the system having an inclination extremely close to edge-on.  The most recent orbital calculation by \cite{2010AJ....140.1623M} showed the eclipses to be highly likely, with a predicted closest projected approach in late January 2015 \citep{2014arXiv1412.1432M}.  As the event approached, three findings showed the prediction likely to be in error.  The source of the mistake was that among the over 600 observations used to determine the orbital model by \cite{2010AJ....140.1623M}, the measurements from 1896.33, 1911.4, and 1937.16 were listed with position angles in error by 180 degrees.  All 3 of these were the last measurements made before closest approach of the binary at their respective epochs, marking a transition from measurements with position angles near 192 degrees to those with 12 degrees or vice-versa.  The stars are nearly equal magnitude, making such mistakes possible.  While efforts were made to find and correct such errors by examining fit residuals (see Figure \ref{fig::114378_resid}, top panel), these three were missed because the orbit model skewed to compensate, as is possible near closest approach, and thus the fit residuals escaped detection (several other 180 degree discrepant measurements were successfully corrected through this method).  The final orbital solution was similarly skewed, which caused errors in the timing.

\section{Catching the Mistake}

Three findings led to identifying these erroneous measurements and the skewed orbit solution.  First, Henrichs and Wijngaarden (preprint, to be submitted) refit the observations obtained from the WDS ignoring all position angle measurements (except to choose a positive or negative sign for the separations) and refitting an orbital model to the separation-only measurements (with inclination set to exactly 90 degrees and ignoring the longitude of the ascending node).  Ignoring the position angle measurements turned out to be a fortunate choice, since the three erroneous position angles no longer skewed the fit.  In a private communication to Muterspaugh on New Year's Eve 2014, they speculated that the eclipse had in fact already passed based on this separation-only orbit evaluation.  However, their orbital solution did not agree well with a full fit including the position angles; seeding a full fit with their parameters led to fitting iterations which eventually converged on the orbital solutions of the \cite{2010AJ....140.1623M} model.  For two weeks it was unclear how to resolve the discrepancy.

Second, in early January 2015 the system was observed by the Navy Precision Optical Interferometer \citep[NPOI, ][]{arm98}.  The visibility trends indicated a binary separation much larger than anticipated by the \cite{2010AJ....140.1623M} orbit model, and it was unclear if the system was still in approach or had already passed conjunction.

Third, a separation measurements by the CHARA array \citep{CHARAEXISTS} on 16 January 2015 found a separation of 45.53 milli-arcseconds (mas) and position angle of 192.85 degrees (which could be 180 degrees ambiguous, though this value is now consistent with our revised orbit below), at a time when the predicted separation from the 2010 orbit was only 7 mas.  After this manuscript was first drafted, two additional measurements were obtained by CHARA and have been included in the orbital solution.  These also show the binary to be growing further apart.  In the CHARA observations, the fringe packet of component ``B'' is about 5\% larger than what has been historically defined by the vast majority of observers as ``A''; this is a function of the stars' differential brightness, diameters, and also may be impacted by the fact that the CHARA measurements were at infrared wavelengths.  To avoid confusion, the historical designations are maintained.

As a result of these findings, the original data set used in \cite{2010AJ....140.1623M} was refit with the uncertainties for position angle artificially increased to absurdly large values (to 10,000 degrees), essentially removing them from the fit, and a model was evaluated using just the separation measurements (with no positive or negative signs) and the PHASES measurements (which were made at a time when the quadrant was unambiguous).  The resulting model was used to calculate predicted position angles for all observation times for comparison with those in \cite{2010AJ....140.1623M}.  It was discovered that the measurements from 1896.33, 1911.4, and 1937.16 were listed with position angles in error by 180 degrees, namely 21.1, 196.4, and 208.3 degrees.  The corrected values are 201.1, 16.4, and 28.3, respectively (see Figure \ref{fig::114378_resid}, second panel).  A new fit was performed with the position angle uncertainties returned to their original values, resulting in an improved $\chi^2$ goodness-of-fit.

\section{Updated Orbit}

The new orbital parameters for $\alpha$ Com were obtained to a combined fit based on the (corrected) non-PHASES, PHASES, and new CHARA array measurements (MJD, $\rho$ (arcseconds), $\theta$ (degrees): 57038.4870368, $0.04553 \pm 0.00081$, $192.85 \pm 1.02$; 57043.4431444, $0.04855 \pm 0.00065$, $193.78\pm 0.77$; 57044.48408989, $0.04947\pm 0.00043$, $193.19\pm 0.50$).  The results are presented in Table \ref{tab::visualOrbits}.  There are 1217 degrees of freedom and $\chi^2 = 1110$.  For comparison with the previous model's epoch of periastron passage, which was listed one full orbit prior, the new model predicts MJD $47615.3 \pm 3.1$ (the increased uncertainty compared to Table \ref{tab::visualOrbits} reflects the uncertainty of the period which impacts this).  The leading indications of the eclipse timing being incorrect are the resulting decreases to both the period and the epoch of periastron passage.

\begin{deluxetable*}{ccccccc}
\tablecolumns{10}
\tablewidth{0pc} 
\tablecaption{Visual Orbit Parameters\label{tab::visualOrbits}}
\tablehead{ 
\colhead{Period (d)} 
& \colhead{${\rm T_\circ}$ (HMJD)} 
& \colhead{Semimajor Axis (arcsec)} 
& \colhead{Eccentricity} 
& \colhead{Inclination (deg)} 
& \colhead{$\omega$ (deg)} 
& \colhead{$\Omega$ (deg)}\\
\colhead{$\sigma_{\rm P}$} 
& \colhead{$\sigma_{\rm T_\circ}$} 
& \colhead{$\sigma_a$} 
& \colhead{$\sigma_e$} 
& \colhead{$\sigma_i$} 
& \colhead{$\sigma_\omega$} 
& \colhead{$\sigma_\Omega$}}\\
\startdata
9443.1 & 57058.37 & 0.67160   & 0.51104  & 90.0576  & 100.578   & 12.215 \\
(3.0) & (0.51)   & (0.00038) & (0.00069) & (0.0094) &  (0.027) &  (0.014) 
\enddata
\tablecomments{
The revised model parameters and fit uncertainties for the binary orbit of $\alpha$ Com.
}
\end{deluxetable*}

\section{New Eclipse Timings for 2014 and Beyond}

Based on the updated orbit, the method outlined in \cite{2014arXiv1412.1432M} was repeated to predict when the system did or will eclipse.  We created 100,000 random sample sets of binary orbit parameters based on the orbital model in Table \ref{tab::visualOrbits}.  In each set, random values were selected for each orbital element using a Gaussian-distributed random number generator with $1\sigma$ width, corresponding to the parameter's formal uncertainty and centered on the best-fit value (e.g.,~values of the period were selected as $9443.1 + 3.0\times g$ days, where $g$ is a standard normal deviate random number).  The resulting set of parameters was then used to calculate the sky-projected separation of the binary every minute from MJD 56955 to 57003 (48 days), to ensure all likely eclipse times were included.  For each set we recorded the time of closest approach, the distance of closest approach, and the duration over which the binary separation was less than 0.7 mas (the approximate diameters of the stars).  If any set failed to produce a minimum separation less than 0.7 mas, it was flagged as non-eclipsing, though in zero trials did this happen.

The most likely time of eclipse was MJD 56981 (20 November 2014, UT).  Unfortunately, the observing season for TSU's photometric observations of $\alpha$ Com did not begin until five days after this predicted time of eclipse due to the star's high air mass near the eastern horizon at sunrise.  The observations were acquired with TSU's T4 0.75m Automatic Photoelectric Telescope (APT).  T4 successfully observed $\alpha$ Com in good conditions for six consecutive nights beginning 25 November 2014, UT.  Those six observations scatter about their mean with a standard deviation of 0.0031 mag and show no evidence for dimming.  APT observations beginning a week or so earlier would have been difficult but perhaps not impossible.

\begin{figure}[]
\epsscale{1.1}
\plotone{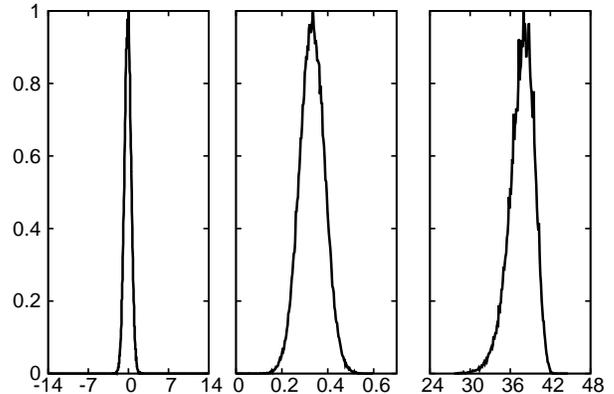}
\caption[HD 114378 ($\alpha$ Com) Eclipse Prediction 1]
{ \label{fig::114378_eclipse}
(Left) Histogram of the time of the eclipse mid-point, versus days since 20 November 2014 (MJD 56981).  (Middle) Histogram of the closest projected separation of the binary (maximum eclipse) in units of milli-arcseconds.  (Right)  Histogram of the duration of the eclipse from ingress at 0.7 mas separation to egress at the same, in units of hours.
}
\end{figure}

\begin{figure}[]
\epsscale{1.1}
\plotone{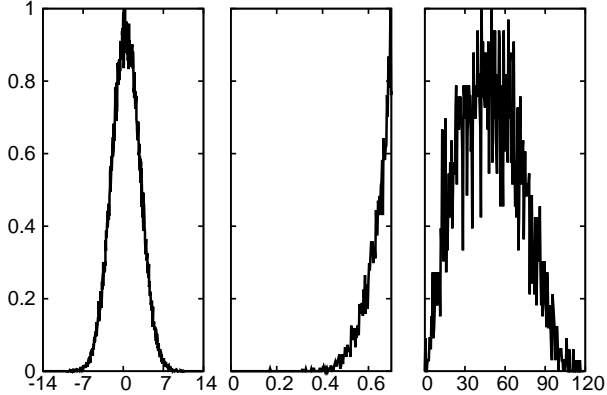}
\caption[HD 114378 ($\alpha$ Com) Eclipse Prediction 2]
{ \label{fig::114378_eclipse2}
Same as as Figure \ref{fig::114378_eclipse} except for the secondary eclipse centered at MJD 61051.
}
\end{figure}

The secondary eclipse is now also a possibility, and is predicted to occur just eleven short years to the week of when the error in the 2014/2015 timing was discovered.  The 100,000 simulations were repeated for a 48 day window beginning on MJD 61032.  The secondary eclipse has only a 3.3\% probability of happening---in only 3318 simulations did the stars come within 0.7 mas of each other.  If the secondary eclipse does occur, the mean predicted time of eclipse is MJD 61051 (11 January 2026); see Figure \ref{fig::114378_eclipse2}.

\begin{figure}[]
\epsscale{1.1}
\plotone{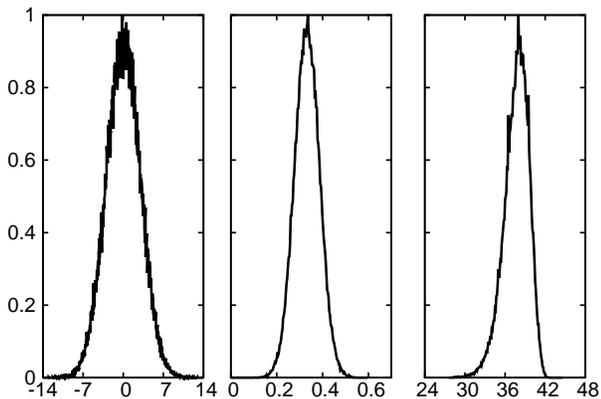}
\caption[HD 114378 ($\alpha$ Com) Eclipse Prediction 3]
{ \label{fig::114378_eclipse3}
Same as as Figure \ref{fig::114378_eclipse} except for the eclipse centered at MJD 66424.
}
\end{figure}

The next primary eclipse is in late September 2040, a time of year which makes this system quite difficult to observe from Earth.  Cameras on distant spacecraft could be used instead.  The 100,000 simulations were repeated for a 48 day window beginning on MJD 66398.  The average time of eclipse is MJD 66424 (27 September 2040); see figure \ref{fig::114378_eclipse3}.

Figures \ref{fig::114378_eclipse}, \ref{fig::114378_eclipse2}, and \ref{fig::114378_eclipse3} show histograms of the results from 100,000 simulated orbital parameter sets for all three events in terms of the time of eclipse minimum, the minimum projected star separation (maximum eclipse), and the duration of the eclipse.

%Figure \ref{fig::EclipseDepth} converts projected sky separation to 
%approximate eclipse depth in magnitudes.

%\begin{figure}[!ht]
%\epsscale{1.0}
%\plotone{mag.eps}
%\caption[Eclipse Depth]
%{ \label{fig::EclipseDepth}
%Depth of eclipse versus projected binary separation; the shallowest simulation (0.55 mas) predicts an eclipse depth of 0.06 mag.  However, only 0.084 percent of trials (84/100,000) predicted a closest approach of 0.5 mas or more, the point at which the eclipse depth exceeds 0.1 mag.
%}
%\end{figure}

\acknowledgements 
We would like to thank and apologize to the many amateur and professional astronomers who were monitoring the system in anticipation of the eclipse.  We thank Chris Farrington, Gail Schaefer, and the CHARA team for providing a timely measurement of the binary separation.  We thank Bob Zavala, Christian Hummel, and the NPOI team for observations helping to lead to these findings.  We thank Brian Mason, Henrique Schmitt, Simon Albrecht, and Francis C.~Fekel for reading early versions of this manuscript.

\bibliography{main}
\bibliographystyle{apj}

\end{document}